\documentclass[twoside]{dis08}
\usepackage[latin1]{inputenc}
\usepackage[dvips]{graphicx,epsfig,color}
\usepackage{wrapfig,rotating}
\usepackage{amssymb,amsmath,array}

\begin{document}
\title{Chiral-Odd Generalized Parton Distributions from Exclusive $\pi^o$ Electroproduction}

\author{Simonetta Liuti$^1$, Gary Goldstein$^2$ and Saeed Ahmad$^3$ 
%
\thanks{U.S. Department
of Energy grants no. DE-FG02-01ER4120 (S.A. and S.L), and 
no. DE-FG02-92ER40702 (G.R.G.)}
%
\vspace{.3cm}\\
%
1,3- University of Virginia - Physics Department \\
382 McCormick Rd., Charlottesville, VA 22904- USA
%
\vspace{.1cm}\\
2- Tufts University - Department of Physics and Astronomy \\
Medford, MA 02155 - USA\\
}

\maketitle

\begin{abstract}
Exclusive $\pi^o$ electroproduction is suggested for extracting both the tensor
charge and the transverse anomalous magnetic moment from experimental data.
A connection between partonic degrees of freedom, given in terms of Generalized Parton
Distributions, and Regge phenomenology is discussed. 
Calculations are performed using a physically motivated 
parametrization that is valid at values of the skewness, $\zeta \neq 0$. 
Our method makes use of information from the nucleon form factor data, 
from deep inelastuc scattering parton distribution functions, and from lattice 
results on the Mellin moments of generalized parton distributions. 
It provides, therefore, a step towards a model independent extraction 
of generalized distributions from the data,
alternative to other mathematical ansatze available in the literature. 
\end{abstract}

\section{Introduction}

Exclusive $\pi^o$ electroproduction from nucleons and nuclei 
allows one to extract the chiral odd Generalized Parton Distributions (GPDs), the 
tensor charge, and other quantities related to transversity,
from experimental data \cite{GGSL1}.   
In this reaction only the C-parity odd combinations
quantum numbers of the t-channel exchanges are, in fact, selected. 
In a description in terms of partonic degrees of freedom, {\it i.e.} based 
on the handbag diagram at leading order, 
the helicity structure for this C-odd process relates to the quark helicity flip, 
or chiral odd generalized parton distributions whose integrals were shown to be related
to the tensor charge $\delta q$, and to the transverse anomalous magnetic moment, $\kappa_q^T$.
This differs markedly from Deeply Virtual Compton Scattering (DVCS), and from both vector meson 
and charged $\pi$ electroproduction, where the axial charge can enter the amplitudes.

In this talk, we start from writing the cross
section and spin asymmetries for  $e p \rightarrow e^\prime \pi^o p$ using the helicity amplitudes formalism.
We then study the sensitivity of the various compoments of the cross section, and of the target transverse
asymmetry, $A_{UT}$, to both $\delta q$ and $\kappa_q^T$, in kinematical ranges where experimental data are
currently being analysed.   
The Compton Form Factor, $\cal{H}$, containing the chiral even GPD, $H$ was already extracted 
with high accuracy from the data, using cross section and bean spin asymmetries measurements.
Within our suggested scenario, the presently analysed $\pi^o$ electroproduction data, as well as future
measurements, will allow one to obtain information on $\cal{H_T}$, and $\cal{\overline{E}_T}$, containing
the chiral-odd GPDs, $H_T$ and $\overline{E}_T = 2 \tilde{H}_T + E_T$. 
We subsequently discuss the structure of our GPDs parametrization in relation with the constraints 
obtained from the data. 

\section{Dependence of $\pi^o$ Electroproduction Observables on Transversity Moments}
The differential cross section for pion electroproduction off an unpolarized target is 
\begin{equation} 
\label{unp-xs}
\frac{d^4\sigma}{d\Omega d\epsilon_2 d\phi dt} = \Gamma \left\{ \frac{d\sigma_T}{dt} + \epsilon_L \frac{d\sigma_L}{dt} 
+ \epsilon \cos 2\phi \frac{d\sigma_{TT}}{dt} 
+ \sqrt{2\epsilon_L(\epsilon+1)} \cos \phi \frac{d\sigma_{LT}}{dt} \right\}.
\end{equation}
where $\epsilon$ is the photon polarization parameter, and for longitudinal polarization alone, 
$\epsilon_L=  (Q^2/\nu^2) \epsilon$; $\Gamma$ is a kinematical factor including 
the Mott cross section,
$\sigma_{M}= 4 \alpha^2 \epsilon_2^2 \cos^2 (\theta/2) /Q^4$.  

The different contributions in Eq.(\ref{unp-xs}) are written in terms of helicity amplitudes, for instance
(we use the notation: $f_{\Lambda_\gamma,\Lambda_N; 0, \Lambda_N\prime}$):
\begin{eqnarray}
\label{dsigT}
\frac{d\sigma_T}{dt} & = & \mathcal{N} \, %
\left( \mid f_{1,+;0,+} \mid^2 + \mid f_{1,+;0,-} \mid^2 + \mid f_{1,-;0,+} \mid^2 +
\mid f_{1,-;0,-} \mid^2 \right)  \nonumber \\
& = & \mathcal{N} \, \left( \mid f_1 \mid^2 + \mid f_2 \mid^2 + \mid f_3 \mid^2 +
\mid f_4 \mid^2 \right)  
\end{eqnarray}
with $ \mathcal{N} = [M (s-M^2)^2]^{-1} \, G$, 
where we used the Hand convention, multiplied by a geometrical factor 
$G=1/8\pi$.
In addition to the unpolarized observables listed above, a number of observables
directly connected to transversity can be written (see {\it e.g.} \cite{GolMor}). 
Here we show the transversely polarized target asymmetry,
\begin{equation}
\label{AUT}
A_{UT} =  2 \Im m (f_1^*f_3 - f_4^*f_2)  \displaystyle \left( d\sigma_T/dt \right).
\end{equation}

It is important to realize that the relations between observables and helicity amplitudes 
is general, independent of any particular model. In \cite{GGSL1} we, 
in fact, used the formalism above in both a Regge model, 
as well as using GPDs.
The connection with the parton model and the transversity distribution is uncovered through the GPD 
decomposition of the helicity amplitudes.
In the factorization scenario the amplitudes for exclusive $\pi^0$ electroproduction, 
$f_{\Lambda_\gamma,\Lambda_N; 0, \Lambda_{N\prime}}$ 
can be decomposed into a ``hard part'', 
$g_{\Lambda_\gamma,\lambda; 0, \lambda^\prime}$ 
and a ``soft part'', $A_{\Lambda^\prime,\lambda^\prime;\Lambda,\lambda}$ 
(where $\Lambda (\Lambda^\prime)$ are the initial (final) proton helicities, and  $\lambda (\lambda^\prime)$
are the initial (final) quark helicities)
through the products of $\gamma^* + q \rightarrow \pi^0 + q$ amplitudes, 
$g_{\Lambda_\gamma,\lambda; 0, \lambda^\prime}$, 
with the matching quark-hadron helicity structures that, in turn, contain the GPDs, in the form

\begin{equation}
f_{\Lambda_\gamma,\Lambda;0,\Lambda^\prime} = \sum_{\lambda,\lambda^\prime} 
g_{\Lambda_\gamma,\lambda;0,\lambda^\prime} 
A_{\Lambda^\prime,\lambda^\prime;\Lambda,\lambda}.
\label{facto}
\end{equation}
The $A_{\Lambda^\prime,\lambda^\prime;\Lambda,\lambda}$ structures are functions of 
$x_{Bj}, t$ and $Q^2$; they are analogous to the Compton Form Factors in DVCS. They 
implicitly contain an integration over unobserved quark momenta. 
Because only one of the $g$ transverse photon functions is non-zero \cite{GGSL1}, 
the relation to the quark-hadron amplitudes is 
quite simple 
{\em Note that because of the pion chirality, {\rm 0$^-$}, the quark must flip helicity at the pion vertex},
\begin{subequations}
\begin{eqnarray}
\label{dsigT2}
\frac{d\sigma_T}{dt} =   \left\{ 
2 \mid A_{++,+-} \mid^2 g_1^2 + \mid A_{--,++} \mid^2 g_2^2  \right. 
\left. + \mid A_{++,--} \mid^2 g_3^2 \right\}   
\end{eqnarray}
\end{subequations}
where we wrote explicitly the $Q^2$-dependence of the $g$ functions from the pion vertex, 
$g_i(\hat{s},t,Q^2) = \hat{g}_i(\hat{s},t)  F_i(Q^2)$, $i=1,5$, for each amplitude. These
depend on whether the $\pi^o$ is produced within an interaction with a vector or an axial
vector meson (see Ref.\cite{GGSL1} for details). 
Similar expressions containing bilinear products of the imaginary and real parts of the 
quark-hadron amplitudes, $A$, can be written for the other contributions.

A formal proof of factorization was given only in the case of longitudinally polarized 
virtual photons producing longitudinally polarized vector mesons
\cite{CFS}.   
Endpoint contributions are surmised to be larger in electroproduction of 
transversely polarized vector mesons, and to therefore prevent factorization.     
Notwithstanding current theoretical approaches, many measurements 
conducted through the years, 
display larger transverse contributions than expected \cite{kubarovsky}.
In this paper we suggest as an alternative avenue a QCD based model, 
that predicts different $Q^2$ behaviors for meson production via natural and unnatural
parity exchanges.

At leading order, using the notations of Ref.\cite{Diehl_01}, one can write the helicity amplitudes
in terms of linear combinations of Meson Production Form Factors (MPFFs).
The MPFFs defining $\pi^o$ production are
\begin{equation}
\mathcal{F} \equiv \mathcal{F}^{p\rightarrow \pi^o} = \frac{1}{\sqrt{2}} \left[ \frac{2}{3} \mathcal{F}^u + 
\frac{1}{3} \mathcal{F}^d \right]
\end{equation}
with $\cal{F} = \cal{H}_T, \cal{E}_T, \widetilde{\cal{H}}_T, \widetilde{\cal{E}}_T$, and the corresponding
$\mathcal{F}^q$ defined in terms of the chiral odd-GPDs, $F\equiv F_T$ as:
\begin{eqnarray}
\mathcal{F}^q(\zeta,t) & = i \pi 
\left[ F^q(\zeta,\zeta,t) - F^{\bar{q}}(\zeta,\zeta,t) \right] +
\mathcal{P} \int\limits_{-1+\zeta}^1 dX \left(\frac{1}{X-\zeta} + \frac{1}{X} \right) F^q(X,\zeta,t).
\label{cal_F}
\end{eqnarray}

%

\section{A Bottom-Up Parametrization of Generalized Parton Distributions}
We performed calculations using a phenomenologically constrained 
model from the parametrization of Refs.\cite{AHLT}.
We summarize the AHLT model for the unpolarized GPD. The parameterization's form is:

\[ H(X,\zeta,t) = G(X,\zeta,t) R(X,\zeta,t), \]

\noindent
where $R(X,\zeta,t)$ is a Regge motivated term that 
describes the low $X$ and $t$ behaviors, while the contribution of 
$G(X,\zeta,t)$, obtained using a spectator model, is centered at intermediate/large values 
of $X$:
\begin{equation}
\label{diq_zeta}
G(X,\zeta,t) = 
{\cal N} \frac{X}{1-X} \int d^2{\bf k}_\perp \frac{\phi(k^2,\lambda)}{D(X,{\bf k}_\perp)}
\frac{\phi({k^{\prime \, 2},\lambda)}}{D(X,{ \bf k}_\perp^\prime)}.   
\end{equation}
Here $k$ and $k^\prime$ are the initial and final quark momenta respectively; explicit expressions
are given in \cite{AHLT}. 
The 
$\zeta=0$ behavior is constrained by enforcing both the forward limit:
$H^q(X,0,0)  =  q_{val}(X)$, 
where $q_{val}(X)$ is the valence quarks distribution, and  
the following relations:
\begin{subequations}
\begin{eqnarray}
\int_0^1 dX H^q(X,\zeta,t)  =  F_1^q(t), \; \; \; \; 
\int_0^1 dX E^q(X,\zeta,t)  =  F_2^q(t),  
\end{eqnarray}
\label{FF}
\end{subequations}
which define the connection with the quark's contribution to the nucleon form factors.
Notice the AHLT parametrization does not make use of a
``profile function'' for the parton distributions, 
but the forward limit, $H(X,0,0) \equiv q(X)$, 
is enforced non trivially. This affords us the flexibility that 
is necessary to model the behavior at $\zeta, \, t \neq 0$. 
$\zeta$-dependent constraints are given by the higher moments of GPDs. 
\begin{wrapfigure}{r}{0.5\columnwidth}
\centerline{\includegraphics[width=0.45\columnwidth]{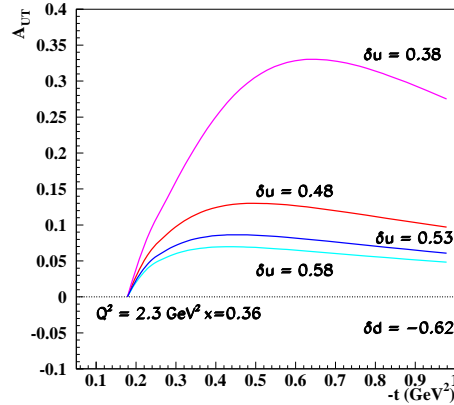}}
\caption{ Transverse spin asymmetry, $A_{UT}$, Eq.(\ref{AUT}). 
}\label{fig1}
\end{wrapfigure}
The $n=1,2,3$ moments of the NS combinations: $H^{u-d} = H^u-H^d$, and $E^{u-d} = E^u-E^d$ 
are available
from lattice QCD \cite{haeg}, $n=1$ corresponding to the nucleon 
form factors. In a recent analysis a parametrization was devised that takes into 
account all of the above constraints. The parametrization gives an excellent description 
of recent Jefferson Lab data in the valence region.  

The connection to the transversity GPDs is carried out similarly to Refs.\cite{Anselmino} for the
forward case by setting:
\begin{eqnarray}
H_T^q(X,\zeta,t) & = & \delta q H^{q, val}(X,\zeta,t) \\
\overline{E}_T^q  \equiv  2 \widetilde{H}_T + E_T & = & \kappa_T^q H_T^q(X,\zeta,t)      
\end{eqnarray}
where $\delta q$ is the tensor charge, and $\kappa_T^q$ is the tensor anomalous
moment  introduced, and connected to the transverse component of the total angular momentum
in \cite{Bur2}. 
Notice that our unpolarized GPD model can be adequately extended to describe $H_T$ 
since it was developed in the valence region, and transversity involves valence quarks only.

In Fig.\ref{fig1} we show the sensitivity of $A_{UT}$ to to the values of the u-quark 
and d-quark tensor charges. The values in the figure 
were taken by varying up to $20 \%$ the values of the tensor charge extracted from the global 
analysis of Ref.\cite{Anselmino}, {\it i.e.}  $\delta u=0.48$ and $\delta d = -0.62$, and fixing the transverse
anomalous magnetic moment values to
$\kappa_T^u = 0.6$ and $\kappa_T^d = 0.3$.
This is the main result of this contribution: it summarizes our 
proposed method for a practical extraction of 
the tensor charge from $\pi^o$ electroproduction experiments. 
Therefore our model can be used to constrain the range of values allowed by the data.

\begin{footnotesize}



%

\end{footnotesize}


\end{document}